\documentclass[reprint, amsmath, amssymb, aps, showkeys, nolongbibliography, numerical]{revtex4-2}

\usepackage[normalem]{ulem}
\usepackage[usenames,dvipsnames]{xcolor}
\usepackage[mathlines]{lineno}

\usepackage{float}
\usepackage{graphicx}
\usepackage[colorlinks=true, linkcolor=blue, allcolors=blue]{hyperref}
\usepackage[mathlines]{lineno}
\usepackage{bm}
\makeatletter
\newsavebox{\@brx}
\newcommand{\llangle}[1][]{\savebox{\@brx}{\(\m@th{#1\langle}\)}%
  \mathopen{\copy\@brx\kern-0.5\wd\@brx\usebox{\@brx}}}
\newcommand{\rrangle}[1][]{\savebox{\@brx}{\(\m@th{#1\rangle}\)}%
  \mathclose{\copy\@brx\kern-0.5\wd\@brx\usebox{\@brx}}}
\makeatother

\begin{document}
\title{Stability of vortex lattices in rotating flows}

\author{Julián Amette Estrada$^{1,2}$}\email{julianamette@df.uba.ar}
\author{Alexandros Alexakis$^{4}$}\email{alexandros.alexakis@phys.ens.fr}
\author{Marc E.~Brachet$^{4}$}\email{brachet@phys.ens.fr}
\author{Pablo D.~Mininni$^{1,2,3}$}\email{mininni@df.uba.ar}
\affiliation{$^1$Universidad de Buenos Aires, Facultad de Ciencias Exactas y Naturales, Departamento de Física, Ciudad Universitaria, 1428 Buenos Aires, Argentina,}
\affiliation{$^2$CONICET - Universidad de Buenos Aires, Instituto de F\'{\i}sica Interdisciplinaria y Aplicada (INFINA), Ciudad Universitaria, 1428 Buenos Aires, Argentina,}
\affiliation{$^3$CNRS-CONICET-UBA, Institut Franco-Argentin de Dynamique des Fluides pour l’Environnement (IFADyFE), IRL2027, Ciudad Universitaria, 1428 Buenos Aires, Argentina,}
\affiliation{$^4$Laboratoire de Physique de l'{E}cole {N}ormale {S}up\'erieure, ENS, Universit\'e PSL, CNRS, Sorbonne Universit\'e, Universit\'e de Paris, F-75005 Paris, France.}

\date{\today}

\begin{abstract}
Vortex lattices---highly ordered arrays of vortices---are known to arise in quantum systems such as type II superconductors and Bose-Einstein condensates. More recently, similar arrangements have been reported in classical rotating fluids. However, the mechanisms governing their formation, stability, and eventual breakdown remain poorly understood. We explore the dynamical stability of vortex lattices in three-dimensional rotating flows. To that end we construct controlled initial conditions consisting of vortex lattices superimposed on turbulent backgrounds. We then characterize their evolution across different 
Rossby numbers and domain geometries.
By introducing an Ekman drag we are able to reach a steady state  where vortex lattices persist with near constant amplitude up until spontaneous breakup of the lattice, or an equivalent of ``melting,'' occurs. We examine an ensemble of runs in order to determine the mean lifetime of the lattice as a function of the system parameters. 
Our results reveal that the stability of the lattices is  a memory-less random process whose mean life-time depends sensitively on the system parameters that if finely tuned can lead to very long lived lattice states. These metastable states exhibit statistical properties reminiscent of critical systems and can offer insight into long-lived vortex patterns observed in planetary atmospheres. 
\end{abstract}

\maketitle

\section{Introduction}

The origin of pattern formation is a central question in physics, in which complex spatial or temporal structures can emerge from the interplay of simple physical laws. Such structures often appear as a result of competing forces, instabilities, or when different time scales in a system reach a dynamical balance. A classical example of this process is provided by Turing patterns \cite{Ouyang_1991}, where the balance between reaction and diffusion processes spontaneously leads to the emergence of stationary chemical patterns. Pattern formation plays a prominent role in a wide range of phenomena, from convective rolls in Rayleigh–Bénard convection \cite{Golubitsky_1984}, the ripples seen in sand dunes \cite{Herrmann_2006}, the formation of ring-like thunderstorms in volcanic eruptions \cite{Ichihara_2023}, the organization of droplets in chiral liquid crystals \cite{Fernandez_2024}, to the formation of large-scale coherent structures in turbulence \cite{Alexakis_2024}. In each case, the emergence of order reflects the system's tendency to self-organize under constraints imposed by geometry, symmetry, and conservation laws.

In statistical mechanics, one of the most commonly observed forms of order is crystalline, in which atoms or molecules arrange into regular periodic structures that minimize the free energy \cite{Ashcroft}. Interestingly, analogous forms of spatial organization can arise in the ordering of topological defects such as vortices. As an example, in type II superconductors, an external magnetic field that penetrates the material with quantized flux lines arranges into a triangular Abrikosov lattice \cite{Abrikosov_1957, Brandt_1986, Severino_2022}. A similar phenomenon takes place in rotating Bose–Einstein condensates, where centrifugal and Coriolis forces and the quantization of circulation lead to the formation of ordered vortex lattices \cite{Fetter_2009}, even in out of equilibrium conditions \cite{Amette_Estrada_2022b}. In all these systems, interacting vortices give rise to crystalline-like patterns governed by long-range interactions and global symmetry constraints \cite{Amette_2025}.

Recently, vortex lattices have been also observed in classical fluids, particularly in systems with broken parity or time-reversal symmetries. As an example, in fluids with odd viscosity, either the modification of the balance of forces by the antisymmetric stress tensor, or a cascade induced process, lead to the emergence of vortex arrays and ordered patterns \cite{Dewit_2024, Floyd_2024}. Vortex lattices have also been observed in rotating convection \cite{Boubnov_1986, Boubnov_1990}, and in instability-driven two-dimensional turbulence \cite{AvK_2024b}. In classical rotating flows a similar ordering has been reported \cite{ClarkDiLeoni2020, Marchetti_2025}. These structures are not just theoretical curiosities: in geophysical and planetary contexts, they can offer insight into the organization of large-scale atmospheric flows. A striking example is the polygonal arrangement of cyclones observed around Jupiter’s poles by the Juno mission \cite{Adriani_2018}, suggesting that the dynamics of rotating fluids can give rise to patterns 
even under highly turbulent conditions. Finally, rotating fluids display an asymmetry between cyclonic and anti-cyclonic motions \cite{Moisy_2010, Gallet_2014} which also manifests in these lattices. Vortex latices have also been found in point vortex models \cite{aref1998point} when only same sign point vortexes are kept indicating that this asymmetry plays an important role, although it still poorly understood.

These previous studies have shown that vortex crystals can form spontaneously in classical fluids and, in particular, in rotating turbulence. However, the mechanisms controlling their stability, the role of the energy balance and turbulent cascades, and the effect of external noise remain unclear. In this study we prepare controlled initial conditions to explore the configuration space of vortex lattices in rotating flows, with the aim of understanding the dynamical stability of the ordered patterns that arise.

\section{The parameter space of rotating turbulence}

We consider an incompressible flow in a three-dimensional (3D) rectangular prism of size $L_x \times L_y \times L_z$, with periodic boundary conditions, in a rotating frame of reference with angular velocity $\boldsymbol{\Omega}$ pointing in the $z$ direction. The equation that determines the dynamics of the fluid is the Navier-Stokes equation, 
\begin{equation}
    \frac{\partial \mathbf{u}}{\partial t} + \mathbf{u} \cdot \boldsymbol{\nabla} \mathbf{u} + 2 \mathbf{\Omega}\times \mathbf{u}=  - \boldsymbol{\nabla} P + \nu \,\Delta {\bf u} - \alpha {\bf u}_\perp^\textrm{2D} + \mathbf{f},
    \label{eq: rotating forced NS}
\end{equation}
where ${\bf u}$ is the velocity field, $\nu$ is the kinematic viscosity, $\alpha$ is an Ekman drag coefficient that may be present or not, ${\bf u}_\perp^\textrm{2D} = (u_x(x,y), u_y(x,y), 0)$ is the two-dimensional (2D) part of the velocity field, $2 \mathbf{\Omega}\times \mathbf{u}$ is the term that accounts for the Coriolis force, and $P$ is a pressure such that incompressibility is satisfied. The external forcing ${\bf f}$ will be assumed to be Gaussian and delta correlated in time, acting only in a narrow Fourier shell with wave vectors $k_f\leq|\mathbf{k}| \leq k_f+\Delta_f$ (where $k_f$ is the forcing wave number), and injecting energy in the system at a rate $\epsilon$. 

We define the mean energy of this system as
\begin{equation}
    E = \frac{1}{2V}\int \left| \mathbf{u}(\mathbf{x}) \right|^2 dV= \frac{1}{2}\sum_{\bf k} \left| \hat{\mathbf{u}}(\mathbf{k}) \right|^2 
\end{equation}
where $\hat{\mathbf{u}}(\mathbf{k})$ is the Fourier transform of $\mathbf{u}(\mathbf{x})$, and the equivalence between the expressions on the r.h.s.~is ensured by Parseval's theorem. Note that we can also decompose the velocity field in Fourier space as
\begin{equation}
    \hat{\bf u}({\bf k}) =
    \begin{cases}
       \hat{\bf u}^\textrm{3D}({\bf k}) & \text {if } k_z \neq 0 , \\
       \hat{\bf u}_\perp^\textrm{2D}({\bf k}_\perp) + \hat{u}_z^\textrm{2D}({\bf k}_\perp) \hat {z} & \text{if } k_z = 0 ,
    \end{cases}   
\end{equation}
where ${\bf k}_\perp = (k_x,k_y,0)$. The Fourier modes with $k_z \neq 0$ are often called 3D, wave, or ``fast'' modes (as they evolve in the time scale of inertial waves), and the Fourier modes with $k_z = 0$ are called two-dimensional (2D) or ``slow'' modes (as the frequency of inertial waves for these modes is identically zero) \cite{Cambon_2004, Sen_2012}. The energy in the 2D modes is then given by
\begin{equation}
    E_\textrm{2D} = \frac{1}{2} \sum_{\bf{k}_\perp} \left[ \left| \hat{\bf u}_\perp^\textrm{2D}({\bf k}_\perp) \right|^2 + \left| \hat{u}_z^\textrm{2D}({\bf k}_\perp) \right|^2 \right] .
\end{equation}

All dimensional quantities in this work are written in units of a unit length $L_0$, a unit velocity $U_0$, and a unit time scale $T_0 = L_0/U_0$. Based on the flow dimensional parameters, we can define several dimensionless numbers. We have the Rossby number at the forcing scale,
\begin{equation}
    \mathrm{Ro} = \frac{\epsilon^{1/3} k_f^{2/3}}{\Omega},    
\end{equation}
the Reynolds number at the forcing scale, 
\begin{equation}
    \mathrm{Re} = \frac{\epsilon^{1/3}}{k_f^{4/3}\nu}.
\end{equation}
and an Ekman-Reynolds number based on the linear drag if drag is present in Eq.~(\ref{eq: rotating forced NS}) as will be done in Sec.~\ref{sec:Ekman},
\begin{equation}
    \mathrm{R}_\alpha = \frac{\epsilon^{1/3}k_f^{2/3}}{\alpha}.
\end{equation} 
Using these quantities we can also define the dimensional eddy turnover time at the forcing scale as $\tau_f=(\epsilon k_f)^{-1/3}$. Finally, based on the domain geometry, we can also define dimensionless aspect ratios of the simulations using the parallel length, 
\begin{equation}
    \lambda_\parallel = \frac{L_z}{\sqrt{L_x L_y}},
\end{equation}
and the perpendicular lengths,
\begin{equation}
    \lambda_\perp = \frac{L_y}{L_x}, 
\end{equation}
while the inverse forcing length scale $k_f$ and the domain height allows us to define
\begin{equation}
    \lambda_f = {k_f L_z} .
\end{equation}

These dimensionless numbers fully describe our system, defining a high dimensional parameter space. 
It is known that in the $\textrm{Re}\to \infty$ limit and large horizontal extend limit $\lambda_\parallel\to 0$, the system can undergo a transition from forward to inverse cascade, depending on the value of $\lambda_f$ and $\textrm{Ro}$ \cite{Sen_2012, Deusebio_2014, vanKan2020, Alexakis2018}.
For large $\textrm{Ro}$ the system becomes independent of rotation,
and the transition appears when a critical value of $\lambda_f$ is crossed \cite{Celani_2010, Benavides_2017, vanKan2020}. For small values of $\textrm{Ro}$ the transition occurs when the product $\textrm{Ro} \, \lambda_f$ crosses a critical value \cite{vanKan2020}. In this parameter space, vortex lattices appear for moderate values of $\textrm{Ro}$ and near the boundary of the transition from forward to inverse cascade \cite{ClarkDiLeoni2020, Marchetti_2025}. Their properties, however, and their dependence on geometric factors such as $\lambda_\perp$ and $\lambda_\parallel$, have not been fully explored.

\begin{table*}
\caption{Simulation parameters: $(L_x,L_y,Lz)/(2 \pi L_0)$ gives the domain sizes in units of $2 \pi L_0$, $(N_x,N_y,N_z)$ is the spatial resolution, $\lambda_\perp$ and $\lambda_\parallel$ are respectively the perpendicular and parallel domain aspect ratios, $\textrm{Ro}$ is the Rossby number, $\mathcal{N}$ is the number of copies done in the cases in which an ensemble of multiple runs was performed, $\textrm{R}_{\alpha^{\ast}}/\textrm{R}_\alpha = \alpha/\alpha^{\ast}$ is the ratio of the Ekman drag coefficient to the optimal value, and $N_v$ is the total number of vortices identified in the domain.}
\label{tab:params}
\centering
\footnotesize
\begin{ruledtabular}
\begin{tabular}{cccccccc}
$(L_x,L_y,L_z)/(2\pi L_0)$ & $(N_x,N_y,N_z)$ & $\lambda_\perp$ & $\lambda_\parallel$ & Ro & $\mathcal{N}$  & $\textrm{R}_{\alpha^{\ast}}/\textrm{R}_\alpha$ & $N_v$ \\ \hline
$(1/2, \sqrt{3}/2, 1)$     & $(256, 432, 512)$ & $\sqrt{3}$   & $1.52$ & $1.15$ & -- & -- & $2$ \\
$(1/2, \sqrt{3}/2, 1)$     & $(256, 432, 512)$ & $\sqrt{3}$   & $1.52$ & $0.87$ & -- & -- & $2$ \\
$(1/2, \sqrt{3}/2, 1)$     & $(256, 432, 512)$ & $\sqrt{3}$   & $1.52$ & $0.70$ & -- & -- & $2$ \\
$(1/2, \sqrt{3}/2, 1)$     & $(256, 432, 512)$ & $\sqrt{3}$   & $1.52$ & $0.58$ & -- & -- & $2$ \\
$(1/2, \sqrt{3}/2, 1)$     & $(256, 432, 512)$ & $\sqrt{3}$   & $1.52$ & $0.50$ & -- & -- & $2$ \\
$(1/2, \sqrt{3}/2, 1)$     & $(256, 432, 512)$ & $\sqrt{3}$   & $1.52$ & $0.44$ & -- & -- & $2$ \\
$(1/2, \sqrt{3}/2, 3/4)$   & $(256, 432, 384)$ & $\sqrt{3}$   & $1.14$ & $1.15$ & -- & -- & $2$ \\
$(1/2, \sqrt{3}/2, 3/4)$   & $(256, 432, 384)$ & $\sqrt{3}$   & $1.14$ & $0.87$ & -- & -- & $2$ \\
$(1/2, \sqrt{3}/2, 3/4)$   & $(256, 432, 384)$ & $\sqrt{3}$   & $1.14$ & $0.70$ & -- & -- & $2$ \\
$(1/2, \sqrt{3}/2, 3/4)$   & $(256, 432, 384)$ & $\sqrt{3}$   & $1.14$ & $0.58$ & -- & -- & $2$ \\
$(1/2, \sqrt{3}/2, 1/2)$   & $(256, 432, 256)$ & $\sqrt{3}$   & $0.76$ & $1.15$ & -- & -- & $2$ \\
$(1/2, \sqrt{3}/2, 1/2)$   & $(256, 432, 256)$ & $\sqrt{3}$   & $0.76$ & $0.87$ & -- & -- & $2$ \\
$(1/2, \sqrt{3}/2, 1/2)$   & $(256, 432, 256)$ & $\sqrt{3}$   & $0.76$ & $0.70$ & -- & -- & $2$ \\
$(1/2, \sqrt{3}/2, 1/2)$   & $(256, 432, 256)$ & $\sqrt{3}$   & $0.76$ & $0.58$ & -- & -- & $2$ \\
$(1/2, \sqrt{3}/2, 3/8)$   & $(256, 432, 192)$ & $\sqrt{3}$   & $0.57$ & $1.15$ & -- & -- & $2$ \\
$(1/2, \sqrt{3}/2, 3/8)$   & $(256, 432, 192)$ & $\sqrt{3}$   & $0.57$ & $0.87$ & -- & -- & $2$ \\
$(1/2, \sqrt{3}/2, 3/8)$   & $(256, 432, 192)$ & $\sqrt{3}$   & $0.57$ & $0.70$ & -- & -- & $2$ \\
$(1/2, \sqrt{3}/2, 3/8)$   & $(256, 432, 192)$ & $\sqrt{3}$   & $0.57$ & $0.58$ & -- & -- & $2$ \\
$(1/2, \sqrt{3}/2, 1/4)$   & $(256, 432, 128)$ & $\sqrt{3}$   & $0.38$ & $1.15$ & -- & -- & $2$ \\
$(1/2, \sqrt{3}/2, 1/4)$   & $(256, 432, 128)$ & $\sqrt{3}$   & $0.38$ & $0.87$ & -- & -- & $2$ \\
$(1/2, \sqrt{3}/2, 1/4)$   & $(256, 432, 128)$ & $\sqrt{3}$   & $0.38$ & $0.70$ & -- & -- & $2$ \\
$(1/2, \sqrt{3}/2, 1/4)$   & $(256, 432, 128)$ & $\sqrt{3}$   & $0.38$ & $0.87$ & $5$ & $1.8$ & $2$ \\
$(1/2, \sqrt{3}/2, 1/4)$   & $(256, 432, 128)$ & $\sqrt{3}$   & $0.38$ & $0.70$ & $5$ & $1.6$ & $2$ \\
$(1/2, \sqrt{3}/2, 1/4)$   & $(256, 432, 128)$ & $\sqrt{3}$   & $0.38$ & $0.70$ & $5$ & $1.4$ & $2$ \\
$(1/2, \sqrt{3}/2, 1/4)$   & $(256, 432, 128)$ & $\sqrt{3}$   & $0.38$ & $0.70$ & $5$ & $1.2$ & $2$ \\
$(1/2, \sqrt{3}/2, 1/4)$   & $(256, 432, 128)$ & $\sqrt{3}$   & $0.38$ & $0.70$ & $5$ & $1.1$ & $2$ \\
$(1/2, \sqrt{3}/2, 1/4)$   & $(256, 432, 128)$ & $\sqrt{3}$   & $0.38$ & $0.70$ & $5$ & $1.0$ & $2$ \\
$(1/2, \sqrt{3}/2, 1/4)$   & $(256, 432, 128)$ & $\sqrt{3}$   & $0.38$ & $0.70$ & $5$ & $0.9$ & $2$ \\
$(1/2, \sqrt{3}/2, 1/4)$   & $(256, 432, 128)$ & $\sqrt{3}$   & $0.38$ & $0.70$ & $5$ & $0.8$ & $2$ \\
$(1/2, \sqrt{3}/2, 1/4)$   & $(256, 432, 128)$ & $\sqrt{3}$   & $0.38$ & $0.70$ & $5$ & $0.6$ & $2$ \\
$(1/2, \sqrt{3}/2, 1/4)$   & $(256, 432, 128)$ & $\sqrt{3}$   & $0.38$ & $0.70$ & $5$ & $0.4$ & $2$ \\
$(1/2, \sqrt{3}/2, 1/4)$   & $(256, 432, 128)$ & $\sqrt{3}$   & $0.38$ & $0.70$ & $5$ & $0.2$ & $2$ \\
$(1, \sqrt{3}/2, 1/4)$     & $(512, 432, 128)$ & $\sqrt{3}/2$ & $0.27$ & $0.87$ & $5$ & $1.0$ & $4$ \\
$(1, \sqrt{3}/2, 1/4)$     & $(512, 432, 128)$ & $\sqrt{3}/2$ & $0.27$ & $0.70$ & $5$ & $1.0$ & $4$ \\
$(1, \sqrt{3}, 1/4)$       & $(512, 864, 128)$ & $\sqrt{3}$   & $0.19$ & $0.87$ & $5$ & $1.0$ & $8$ \\
$(1, \sqrt{3}, 1/4)$       & $(512, 864, 128)$ & $\sqrt{3}$   & $0.19$ & $0.70$ & $5$ & $1.0$ & $8$ \\
$(2, \sqrt{3}, 1/4)$       & $(1024, 864, 128)$& $\sqrt{3}/2$ & $0.13$ & $0.87$ & $5$ & $1.0$ & $16$ \\
$(2, \sqrt{3}, 1/4)$       & $(1024, 864, 128)$& $\sqrt{3}/2$ & $0.13$ & $0.70$ & $5$ & $1.0$ & $16$ \\
$(4, 2\sqrt{3}, 1/4)$      & $(2048, 1728, 128)$ & $\sqrt{3}/2$& $0.067$& $0.87$ & $5$ & $1.0$ & $64$ \\
\end{tabular}
\end{ruledtabular}
\end{table*}

\section{Preparation of vortex lattices}

The parameter space previously described is a key component of the question that we want to unravel in this work. Here we study coherent vortex structures, similar to those previously seen in \cite{ClarkDiLeoni2020, Marchetti_2025}. However, in those cases, the vortex lattices were allowed to develop spontaneously from the system dynamics. This results in the generation of imperfect lattices, or, as a result of the aspect ratios chosen for the domains, in lattices that cannot satisfy the boundary conditions and must have defects. As we are interested in identifying conditions and regions of parameter space in which the lattices remain dynamically stable, we need to prepare states that reduce any imperfection.

The possible geometries that vortex lattices can develop were studied in detail in \cite{Aref_2003}. To study different vortex lattice configurations and their stability, we generate initial conditions in which such structures are already embedded. This is achieved through a two-step procedure. First, we run a simulation solving Eq.~(\ref{eq: rotating forced NS}), with the same physical parameters as the target case, applying a three-dimensional random forcing ${\bf f}$. After a sufficiently long evolution---during which large-scale structures may emerge---we extract a velocity field snapshot, and remove all 2D slow modes, thereby retaining only the 3D modes, $\bf{u}^\textrm{3D}$. Second, we construct a velocity field composed of an ideal lattice of same-sign vortices embedded in 2D slow modes. This is done with a stream-function designed to place vortices at prescribed positions $\{(x_1, y_1), (x_2, y_2), \dots, (x_n, y_n)\}$, consistent with the geometry and boundary conditions of the simulation domain. The stream-function is given by
\begin{eqnarray}
    \psi(\mathbf{x}) = \sum_{i=1}^{n} A \exp \left\{ \frac{1}{\beta^2} \left[ \cos\left(\frac{x - x_i}{L_x}\right) + \nonumber \right. \right. \\ 
    \left. \left. \left(\frac{L_y}{L_x}\right)^2 \cos\left(\frac{y - y_i}{L_y}\right) -\frac{L_x^2+L_y^2}{L_x^2} \right] \right\} ,
\end{eqnarray}
where $A$ sets the amplitude of the vortices, and $\beta$ controls their width. Note that close to ($x_i,y_i)$ $\psi(\mathbf{x})$, has a gaussian profile.
The 2D velocity field associated with this stream-function is defined as ${\bf u}_\perp^\textrm{2D} = (\partial_y \psi, -\partial_x \psi, 0)$. 
The total initial velocity field is finally constructed as
\begin{equation}
    \mathbf{u}_0({\bf x}, t=0) = \mathbf{u}^\textrm{3D}({\bf x}) + \mathbf{u}_\perp^\textrm{2D}({\bf x}),
\end{equation}
and it depends explicitly on the number and arrangement of vortices prescribed in the streamfunction.

\begin{figure}
    \centering
    \includegraphics[width=\linewidth]{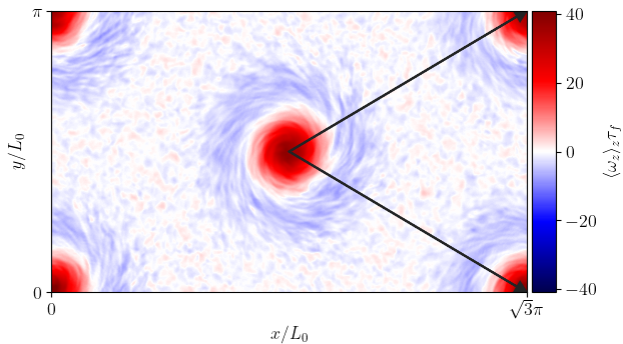}
    \caption{The simplest triangular lattice in a domain with aspect ratio $\lambda_\perp = 1/\sqrt{3}$. The colors show $\left< \omega_z \right>_z \tau_f$ (i.e., the vertically averaged vertical vorticity in units of $\tau_f^{-1}$) in a simulation with $\textrm{Ro} = 0.70$ and $(L_x,L_y,L_z)=(1/2,\sqrt{3}/2,1)2\pi L_0$. The Bravais lattice vectors are indicated by the black arrows.}
    \label{fig:wz_sample}
\end{figure}

We consider triangular vortex lattices compatible with periodic domains \cite{Abrikosov_1957, Marchetti_2025}. A triangular lattice is one of the five 2D Bravais lattices, characterized by primitive vectors forming an angle of $\pi/3$. 
As the computational domain is rectangular with horizontal dimensions $(L_x,L_y)$, only certain geometric factors $\lambda_\perp$ are compatible with triangular lattice without defects, given by $(L_x,L_y)=a (\sqrt{3}\,n , m)$ where $n,m$ are integers and $a$ is the nearest-neighbour distance between vortices. 
The smallest rectangular domain compatible with a triangular lattice is therefore given by $a(\sqrt{3} , 1)$ with $L_y = a$. This rectangular cell contains two vortices; without loss of generality we can place one at the centre of the domain and the other at one of the corners, taking periodicity into account (see Fig.~\ref{fig:wz_sample} for an example). 
We note that both vortices in Fig.~\ref{fig:wz_sample} have the same sign of vorticity, which makes them different from the classical vortex dipole observed in 2D flows.  
Larger lattices with the same inter-vortex distance can be constructed by choosing larger values of $(n,m)$, such that $L_x=\sqrt{3} \,na$ and $L_y= ma$. 

\section{Numerical  methodology}

As we aim at studying the stability of these vortex lattices, once we have the initial states, we integrate them numerically for long times and study their potential deformation and break up time. This will be done in two steps. 
In the first step, discussed in Sec.~\ref{sec:free}, we consider forced states without drag, i.e., we integrate numerically the initial states using Eq.~(\ref{eq: rotating forced NS}) with $\alpha = 0$ while we keep applying the random forcing ${\bf f}$. 
In the second step, considered in Sec.~\ref{sec:Ekman}, we look for ways to generate steady states that survive for arbitrarily long times, by balancing energy injection with large-scale friction. To this end we integrate Eq.~(\ref{eq: rotating forced NS}) with Ekman drag ($\alpha \neq 0$) while keeping the random forcing. The complete set of all simulations performed, with their parameters, is summarized in table \ref{tab:params}.

All simulations are direct numerical simulations (DNSs), i.e., they resolve explicitly all relevant time and length scales. A parallel pseudo-spectral code with the $2/3$ rule for dealiasing is used to integrate Eq.~(\ref{eq: rotating forced NS}); the code, GHOST, is publicly available \cite{Mininni2011, Rosenberg_2020}. In all simulations, $k_f = 20/L_0$, and the spectral forcing width is $\Delta_f = 2/L_0$. We use spatial resolutions with $\Delta x \approx \Delta y \approx \Delta z \approx 1.2 \times 10^{-2} L_0$, a condition that is kept fixed for all box sizes and aspect ratios. As a result, the maximum resolved wave number after the $2/3$ dealiasing rule is $k_\mathrm{max} \approx 170/L_0$. All simulations have the same viscosity, such that $\mathrm{Re} \approx 13$ within $<10\%$ variations; even though this Reynolds number is small, note that it is evaluated at the forcing scale, and limited by spatial resolution. The resulting Kolmogorov dissipation scale, $k_{\eta}= (\epsilon/\nu^3)^{1/4}$, is smaller than $k_{\mathrm{max}}$ in all cases.

\section{Free large-scale evolution\label{sec:free}}

As already mentioned, we fist consider cases with random forcing but without Ekman drag, in such a way that the large scales can evolve freely. We focus on the smallest lattice allowed in a given domain, with $N_v=2$ vortices as shown in Fig.~\ref{fig:wz_sample}, varying the domain height and $\textrm{Ro}$ (see table \ref{tab:params}). 

\subsection{Fixed height dynamics}

We consider first the simulations with fixed height and with different Rossby numbers. To this end, we focus on the simulations in table \ref{tab:params} with box size $(L_x,L_y,L_z)=(1/2,\sqrt{3}/2,1)2\pi L_0$ and with $\mathrm{Ro} \in [0.44, 1.15]$.

\begin{figure}
    \centering
    \includegraphics[width=\linewidth]{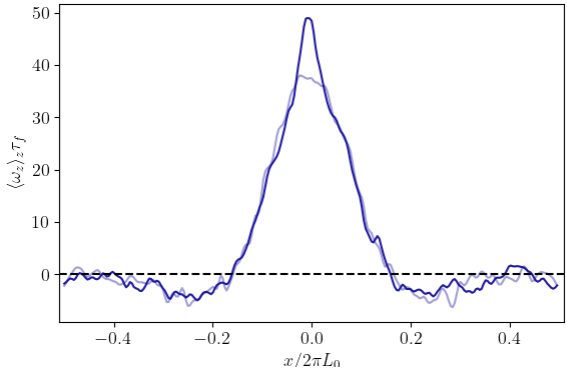}
    \caption{Vertically averaged vertical vorticity, $\left< \omega_z \right>_z$, in units of $\tau_f^{-1}$, as a function of $x$ and for a value of $y$ that goes across the center of a vortex in a simulation with $\textrm{Ro} = 0.70$, $\lambda_\perp=\sqrt{3}$ and $\lambda_\parallel = 1.52$. Light blue shows the initial condition, and dark blue corresponds to a later time.}
    \label{fig:vortex_profile}
\end{figure}

As time evolves the system tends to remain in the ordered configuration shown in Fig.~\ref{fig:wz_sample}, without significant changes in the vortex positions. However, the detailed profile of the vortices changes from the initial conditions, developing a sharper structure in the maximum of vorticity as shown in Fig.~\ref{fig:vortex_profile}. The figure shows the vertically averaged profile of the vertical vorticity, $\langle \omega_z\rangle_z$, as a function of $x$, and for a value of $y$ that goes across the center of a vortex, at the initial time and at a later time. All vortices have similar shapes and amplitudes, with large positive values of $\omega_z$ at their centers, surrounded by a ring of small negative vorticity that decays with the distance from the vortex center (see also Fig.~\ref{fig:wz_sample}). We note that the integral of the vorticity over the entire region is zero as  in the periodic domain the flow total vorticity must be zero.

\begin{figure}
    \centering
    \includegraphics[width=1\linewidth]{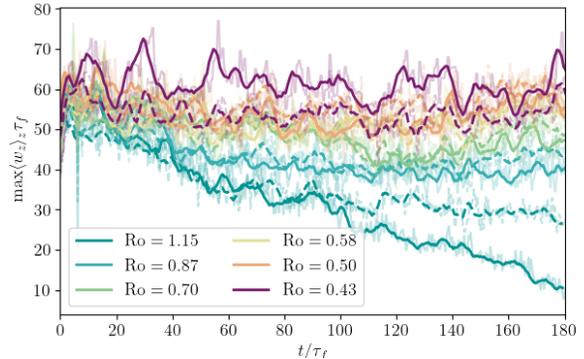}
    \caption{Peak vertical vorticity as a function of time of the two vortices in simulations with $N_x=256$, $N_y=432$, $N_z=512$, and with different values of $\textrm{Ro}$ from $0.43$ to $1.15$. For each color, corresponding to a value of $\textrm{Ro}$, the light solid and dashed lines indicate the instantaneous value of $\max \{\omega_z\}$ in each vortex, while the dark lines give the smoothed evolution.}
    \label{fig:amplitude evolution}
\end{figure}

After this transient, the vertical vorticity in each vortex can either slowly decrease or remain approximately constant. This can be seen in Fig.~\ref{fig:amplitude evolution}, which shows the maximum of $\omega_z$ for each of the two vortices in all simulations with $N_x=256$, $N_y=432$, and $N_z=512$.  
Despite the fact that vortices are far apart and somewhat shielded by the ring of opposite sign vorticity, their amplitudes evolve in similar ways, indicating the system remains correlated at large scales.
This is true for all simulations except for the case with $\textrm{Ro} > 1$. In this case, $\max \{\omega_z\}$ decreases in time, and for $t \gtrsim 90 \tau_f$ the evolution of the two vortices drifts apart in amplitude with one vortex decaying faster than the other. In this simulation, the time evolution results in the disappearance of the vortices at late times. 

\begin{figure}
    \centering
    \includegraphics[width=1\linewidth]{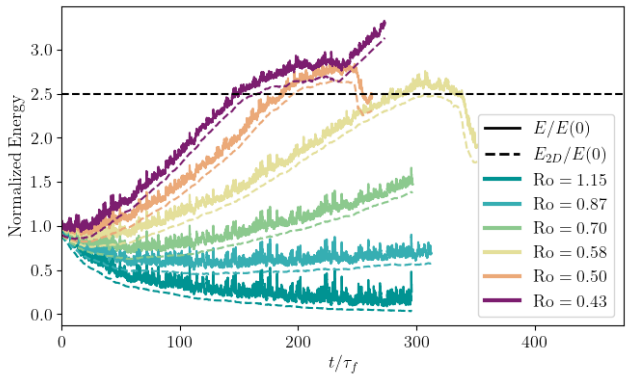}
    \caption{Total (solid lines) and 2D (dashed lines) energies in the simulations with size $(L_x,L_y,L_z)=(1/2,\sqrt{3}/2,1)2\pi L_0$, with values of $\textrm{Ro}$ from $0.43$ to $1.15$. The energies are normalized by the initial value. A value $E/E(0)=2.5$ is marked by a dashed black line as a reference.}
    \label{fig:energy_evolution}
\end{figure}

We now look at the time evolution of the total and 2D energy components. 
The amount of energy in the 2D modes relative to the total energy gives a first quantification of how two-dimensional the flows are. 
The evolution of these quantities in all simulations with $N_x=256$, $N_y=432$, and $N_z=512$ is shown in Fig.~\ref{fig:energy_evolution}. 
In all cases, the difference between $E$ and $E_\textrm{2D}$ remains approximately constant in time. 
This indicates that the 3D energy $E_\textrm{3D}=E - E_\textrm{2D}$ does not grow in time and any growth or decay in $E$ is due to $E_\textrm{2D}$ i.e. the vortex latice amplitude. 

The evolution of $E$ in Fig.~\ref{fig:energy_evolution} is sensitive to the Rossby number. The simulation with $\textrm{Ro}=1.15$ shows a decay of $E_\textrm{2D}$ with time, while all other simulations, with smaller values of $\textrm{Ro}$, display a growth of energy with time after a short transient. 
In real space, these differences manifest as contrasting dynamics: 
At large $\textrm{Ro}$ and as already mentioned, the vortices eventually decay and the lattice disappears (see Fig.~\ref{fig:amplitude evolution}). This suggests that in this case, rotation is insufficient to generate an inverse cascade and stabilize the large-scale 2D structures. 
At intermediate values of $\textrm{Ro}$ the energy grows in time slowly, and for as long as we integrate the system, we observe a lattice.
And finally, for smaller values of $\textrm{Ro}$ (i.e., in the simulations with $\textrm{Ro}=0.58$ or smaller), energy grows until a time at which the vortices accumulate too much energy in 2D modes, and the lattice eventually breaks up, due to merging of the two vortices.
This results in a saturation in the growth of $E$ and $E_\textrm{2D}$, followed by a decay in some of the runs, and by a new growth in other cases as columnar vortices are recreated. 
It is worth noting that the energy amplitude that this transition happens is approximately the same
for all $Ro$ that reached this amplitude.
In Fig.~\ref{fig:energy_evolution} we indicate only as a reference a value of the energy equal to $2.5 E(0)$, where $E(0)$ is the energy in the initial conditions. 
Interestingly, the instability of individual columnar vortices as energy grows in rotating flows has been recently studied in \cite{seshasayanan2020onset,lohani2024effect,Das_2025}. 

Our simulations thus indicate that, even after carefully preparing initial ordered states, the vortex lattice does not settle into a stationary state. It is either gaining energy when $\textrm{Ro}$ is too high and breaking the lattice due to vortex merging when an energy threshold is crossed or it is losing energy when $\textrm{Ro}$ is too low making the 2D vortices disappear into 3D turbulence. This suggests that only a specific, fine-tuned value of $\textrm{Ro}$ can result in a vanishing time derivative of the energy for which a stationary state of the vortex lattice is reached.  
Thus, the set of stationary cases must have zero measure in parameter space.

\subsection{Finite-time stability phase space}

\begin{figure}
    \centering    
    \includegraphics[width=1\linewidth]{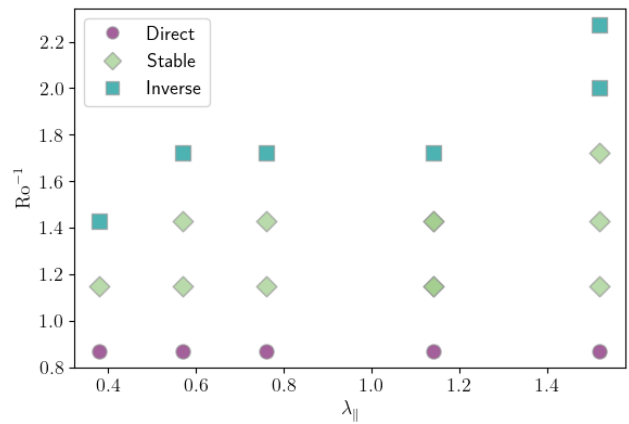}
    \caption{Phase space in terms of $\lambda_\parallel$ and $\mathrm{Ro}^{-1}$, where the different colored markers represent the end state of the simulation after a time $T/\tau_f = 273$. Points marked as ``direct'' correspond to cases in which $E$ decays, points marked as ``inverse'' correspond to cases in which $E$ grows and the lattice breaks up, while ``stable'' indicates cases in which at time $T$ the lattice was still visible in the simulations.}
    \label{fig:phase space}
\end{figure}

Based on the previous discussion we can only explore the presence of the lattices only for a finite observation time. To construct this finite-time phase space, we characterize the possible outcomes of simulations at a given time $T$ while varying two key control parameters: the height of the domain measured by $\lambda_\parallel$, and the Rossby number. Three types of long-term outcomes are identified: ``vortex lattice states'' (i.e., cases in which the vortices are still present at time $T$, and ordered according to their initial positions or rigidly shifted in space), ``direct-cascade'' cases (i.e., cases in which the vortices dissolve, and the energies $E$ and $E_\textrm{2D}$ only decay in time, indicating that the injected energy is transferred to smaller scales where it is dissipated), and ``inverse-cascade'' cases (i.e., cases in which $E$ and $E_\textrm{2D}$ grow in time as a result of energy upscaling, and the lattice breaks up when vortices become too energetic or when vortices merge as the inverse energy cascade proceeds to larger scales). 

The resulting phase space is shown in Fig.~\ref{fig:phase space}, constructed for an observation time fixed at $T = 273 \tau_f$.  This time greatly exceeds the characteristic dynamical timescales of all runs, as it is much larger than the eddy turnover time at the forcing scale, and even larger than the time defined by the inverse of the maximum of vorticity, $1/\max\{\omega_z\} \approx 0.02 {\tau_f}$. The results show a clearly delineated region of finite-time stability, flanked by zones in which the fate of the lattice is decided by the dominance of either the inverse or the direct energy cascade.

As already mentioned, while this analysis provides a partial picture of the lattice persistence, it is limited by the fixed observation time. To further quantify the intrinsic dynamical stability of the vortex lattice, we need to prepare stationary energy states. This will be considered in the next section.

\section{Controlled energy states \label{sec:Ekman}}

\begin{figure}
    \centering    \includegraphics[width=1\linewidth]{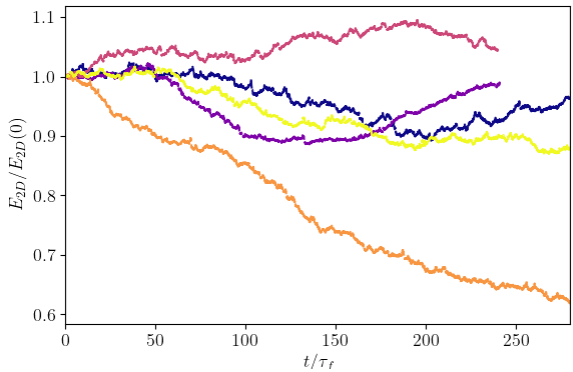}
    \caption{Time evolution of $E_\textrm{2D}$ normalized by its initial value, for an ensemble of simulations with the same initial condition and in a domain of size $(L_x,L_y,L_z)=(1/2,\sqrt{3}/2,1/4)2\pi L_0$ with $\mathrm{Ro}= 0.87$ and $\textrm{R}_{\alpha^*}=\textrm{R}_\alpha=3710$ (i.e., $\textrm{R}_{\alpha^*}/\textrm{R}_\alpha = 1$).}
    \label{fig:multiple instances evolution}
\end{figure}

\subsection{The procedure to adjust the Ekman drag}

In many experimental and natural settings, rotating flows are influenced by boundary layers which alter the energy balance, and can result in statistically stationary states even in the presence of an inverse energy cascade as a result of its balance with a large-scale friction. Considering this as a motivation, in this section we consider situations in which Ekman drag is present, i.e., $\alpha \neq 0$ in Eq.~(\ref{eq: rotating forced NS}). 
The selection of the drag coefficient $\alpha$ is difficult, as it has a complex interaction with the system dynamics, and directly competes with the condensation of energy in low Rossby cases. We then consider it as a Lagrange multiplier that can be used to impose a mean value for the energy in the system steady state. To this end we couple Eq.~(\ref{eq: rotating forced NS}) to a control equation for $\alpha$ that dynamically adjusts the damping strength in order to maintain a prescribed energy $E_\textrm{2D}^*$ in the slow modes. Specifically, the evolution of $\alpha$ satisfies
\begin{equation}
    \frac{\partial \alpha}{\partial t} = \gamma (E_\textrm{2D} - E_\textrm{2D}^*),
    \label{eq:Lagrange}
\end{equation}
where $\gamma$ sets a relaxation timescale.

We integrate Eqs.~(\ref{eq: rotating forced NS}) and (\ref{eq:Lagrange}) in the presence of a lattice and for a given set of parameters, until the system reaches a steady state around the target energy, and we denote $\alpha^* = \left< \alpha \right>_t$ as the time-averaged value of $\alpha$ in this steady state. The final state of these simulations will be used as the initial condition of subsequent runs. We also use $\alpha^*$ as the reference value to perform the subsequent simulations with fixed values of the Ekman drag coefficient, i.e., solving only Eq.~(\ref{eq: rotating forced NS}) with constant $\alpha$.

\begin{figure}
    \centering
    \includegraphics[width=0.9\linewidth]{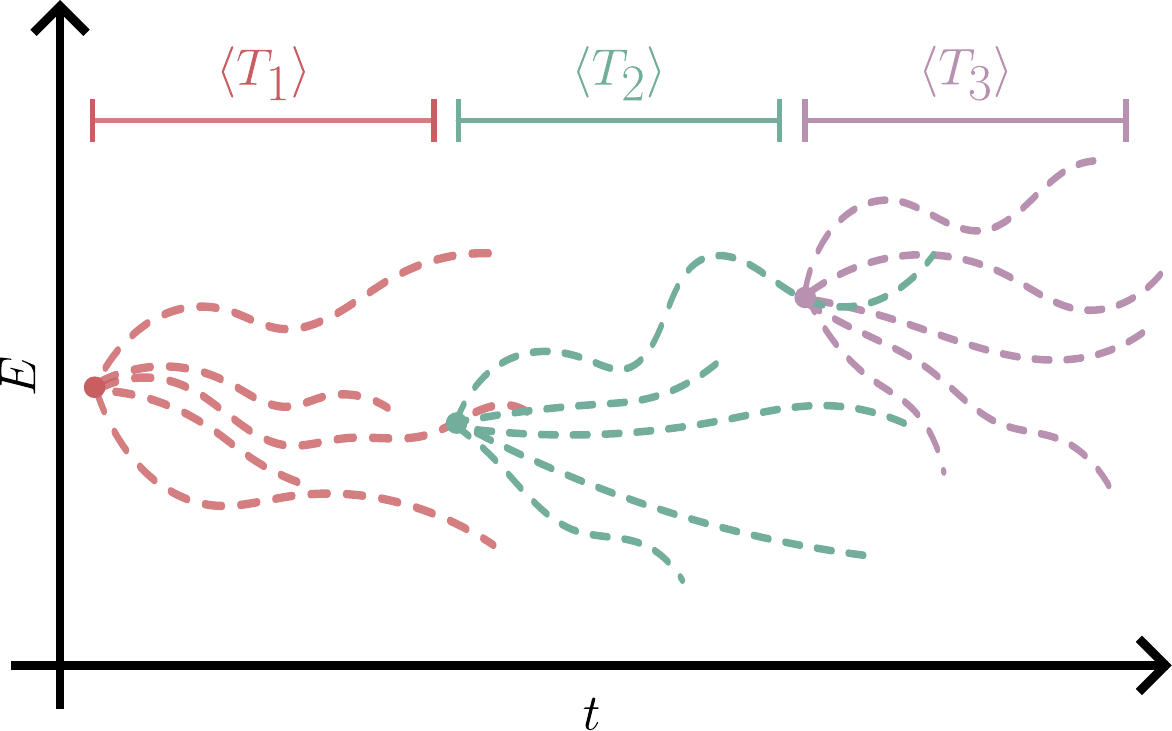}
    \caption{Sketch of the procedure used to study the system's memory of the initial state. An ensemble of five runs is done (red), and a prior time of the lattice that remained stable at the latest time is picked as the initial condition for a new ensemble of five runs. This process is repeated multiple times.}
    \label{fig:restarts}
\end{figure}

\subsection{The need for ensembles}

We first consider  simulations in a domain of size $(L_x,L_y,L_z)=(1/2,\sqrt{3}/2,1/4)2\pi L_0$
and with $\mathrm{Ro}= 0.87$ and $\textrm{R}_\alpha = \textrm{R}_\alpha^* = 3710$ (i.e., with $\alpha = \alpha^*$). 
An ensamble of runs is performed where the same initial lattice is used as initial condition and this configuration is perturbed only by different realizations of the small-scale random forcing, all sharing the same properties except for the random phases. Figure \ref{fig:multiple instances evolution} shows the time evolution of $E_\textrm{2D}$ in these runs, illustrating how the outcome varies significantly across the different realizations with the same parameters. Moreover, while some lattices in this ensemble destabilize quickly, others remain coherent and stable over long times. This variability confirms the stochastic nature of the system, even though the forcing acts at scales ten times smaller than the lattice spacing.

\begin{figure}
    \centering
    \includegraphics[width=1\linewidth]{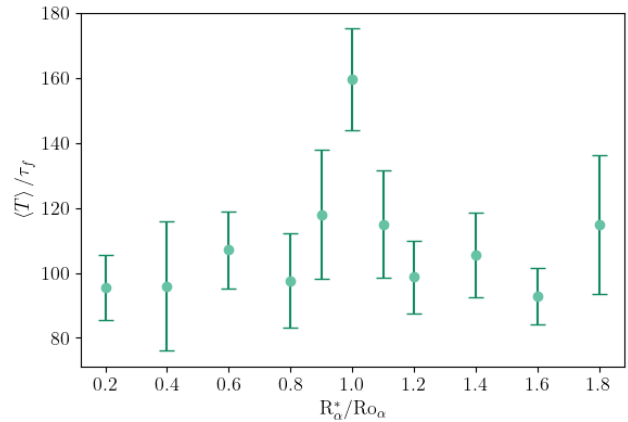}
    \caption{Mean lifetime of the lattice, $T_\textrm{latt}$, in ensembles of simulations with different values of $\textrm{R}_{\alpha^*} / \textrm{R}_\alpha$, and for cases with $\mathrm{Ro}=0.70$ and domain size $(L_x,L_y,L_z)=(1/2,\sqrt{3}/2,1/4)2\pi L_0$. The error bars indicate the standard error, $\sigma_{T_\textrm{latt}}/\sqrt{\mathcal{N}}$, where $\sigma_{T_\textrm{latt}}$ is the standard deviation in all values of $T_\textrm{latt}$, and $\mathcal{N}$ is the number of runs in the ensemble.}
    \label{fig:tunning}
\end{figure}

To show that this is a memory-less process and does not depend on the exact choice of initial conditions we conducted a series of ensembles of runs with $(L_x,L_y,L_z)=(1/2,\sqrt{3}/2,1/4)2\pi L_0$, $\mathrm{Ro}= 0.7$, and $\textrm{R}_\alpha = 0.83 \, \textrm{R}_\alpha^*$ 
The ensemble of runs is started following the same procedure as the one shown in Fig.~\ref{fig:multiple instances evolution}: five simulations are started from the same initial condition, with different random sequences of phases for the forcing. Then, we pick a simulation that still shows a lattice at the last time, and pick a prior time (to ensure no breaking process has started) as the initial condition for a new ensemble of five runs. This process is repeated once again (see Fig.~\ref{fig:restarts}). We found that the statistical distributions of the lattice lifetimes in the three consecutive ensembles are indistinguishable from each other, confirming that the memory of the initial conditions is lost. This is consistent with the notion of metastable states: the lattice represents a local energy minimum from which the system departs only when perturbations are strong enough to overcome a certain energy barrier. This behavior is the same for the different values of $\textrm{R}_\alpha$ explored.
It is worth mentioning here that this property distinguishes these vortex lattices from those seen, e.g., in superconductors or in superfluids. In those cases, both experimentally and numerically, large resistance of the lattice to fluctuations has been reported \cite{Yarmchuk79,Kragset2006, Amette_2025}.

\subsection{The lattice lifetime as a random variable}

From the previous observations, we must interpret the measured lattice lifetimes as realizations of a random variable and only measurements of means over ensembles of runs are of value.

To study how the mean lattice lifetime is affected by the Ekman drag, we consider the ensembles with $(L_x,L_y,L_z)=(1/2,\sqrt{3}/2,1/4)2\pi L_0$, $\mathrm{Ro}=0.7$, and with different values of $\textrm{R}_\alpha$. For $\textrm{R}_\alpha \neq \textrm{R}_\alpha^*$, a non-zero derivative of $E$ and $E_\textrm{2D}$ develops in time. For each value of $\textrm{R}_\alpha$ we measured the lifetime of the lattice in each run in the ensemble. Figure \ref{fig:tunning} shows the result, with the mean value of the lattice lifetime $\langle T_\textrm{latt} \rangle$, and its deviation in the ensemble. The lifetime reaches a maximum for $\textrm{R}_\alpha/\textrm{R}_\alpha^* = 1$, the case in which the system's energy remains approximately constant. As $\textrm{R}_\alpha$ deviates from this value---either increasing or decreasing---the lifetime drops sharply and approaches a plateau. Remarkably, the mean lifetime in this plateau appears largely independent of whether the system gains or loses energy, indicating that the instability is not directly tied to the direction of the energy cascade, but rather to the magnitude of the energy imbalance.

This indicates that when the system is energetically out of equilibrium, fluctuations tend to draw energy into or out of the lattice structure, destabilizing it. 
We thus conclude that, for a given lattice configuration, domain size, and Rossby number, there exists an optimal $\textrm{R}_\alpha$ that maximizes the lattice lifetime on average. This condition corresponds to a dynamically stable balance between the lattice structure and the energetic flow environment.

\begin{figure}
    \centering
    \includegraphics[width=1\linewidth]{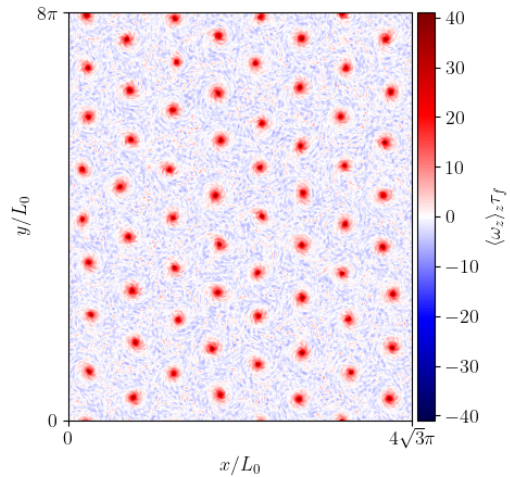}
    \caption{Vertically averaged vertical vorticity, $\left< \omega_z \right>_z$, in a simulation with $\mathrm{Ro}=0.87$ and $\textrm{R}_\alpha^* = 3710$ at $t= 82 \tau_f$, in a domain with $L_z=\pi/4$ and $(L_x,L_y)= \pi(n, m \sqrt{3})$ with $(m,n)=(8,4)$.}
    \label{fig:64 vortices}
\end{figure}

\begin{figure}
    \centering
    \includegraphics[width=1\linewidth]{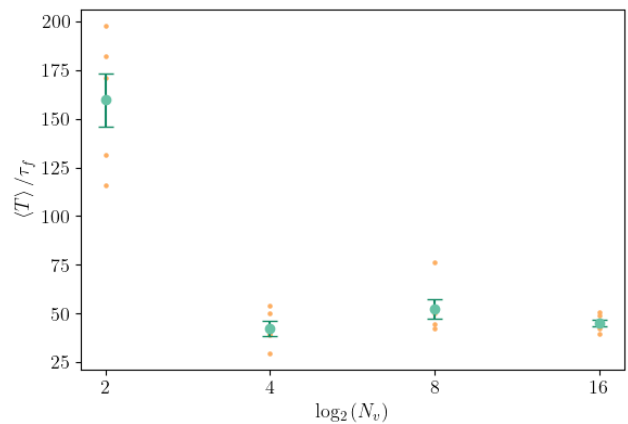}
    \caption{Mean lifetime (green dots) of simulations with lattices with different number of vortices, $N_v$. The error bars represent the standard error in green, while the actual value of $T_\textrm{latt}$ measured in each individual run is shown as orange dots. The results correspond to runs with $\mathrm{Ro}=0.7$, $\textrm{R}_\alpha = \textrm{R}_\alpha^* = 3184$, $L_z=\pi/4$, and $(L_x,L_y)= \pi(n, m \sqrt{3})$ with $(m,n)\in \left\{(1,1), (2,1), (2,2), (4,2)\right\}$. }
    \label{fig:lifetime}
\end{figure}

\subsection{The effect of the number of vortices}

So far we examined boxes of minimal size. The stability of a lattice however can depend on its size. We thus explore finally, how lattice stability is affected by its size {\it ie} by the number of vortices present in the initial condition. To this end, we constructed larger vortex lattices by tessellating space, using periodic replicas of the two-vortex unit cell. An example of such a construction is shown in  Fig.~\ref{fig:64 vortices}.
For each of these configuration, we run ensembles of five simulations using $\mathrm{Ro}=0.7$ and $\textrm{R}_\alpha = \textrm{R}_\alpha^* = 3184$. In these simulations $L_z=\pi/4$ and $(L_x,L_y)= \pi(n, m \sqrt{3})$ with $(m,n)\in \left\{(1,1), (2,1), (2,2), (4,2) \right\}$ (see table \ref{tab:params}).

The result of measuring the lifetime of these lattices is shown in Fig.~\ref{fig:lifetime}.  The smallest lattice (two vortices) exhibits significantly longer lifetimes than all other configurations with a larger number of vortices. However, variance in the lifetimes decreases with the number of vortices in the lattice, suggesting that larger lattices exhibit more predictable albeit shorter-lived dynamics. The presence of a plateau in the lifetime also signals that the system as it becomes larger its stability is independent of the number of vortices.

The numerical cost of these large vortex lattices prohibited us from further exploring different parameter values. 
We mention here however that we repeated the same procedure for a different set of parameters ($\mathrm{Ro}=0.87$, $\textrm{R}_\alpha = \textrm{R}_\alpha^* = 3710$, closer to the expected stable region). In that case, fluctuations in the lifetime were large in all cases, with some simulations that did not destabilize even after very long integration times $T\ge 100 \tau_f$. 
 
A good example of this regime of very large and long-lived arrays is a case with 64 vortices with $(m,n)=(8,4)$, shown in Fig.~\ref{fig:64 vortices}. In this case, the vortex lattice remained coherent for at least $T = 105 \tau_f$, with no signs of destabilization.

\begin{figure}
    \centering
    \includegraphics[width=0.8\linewidth]{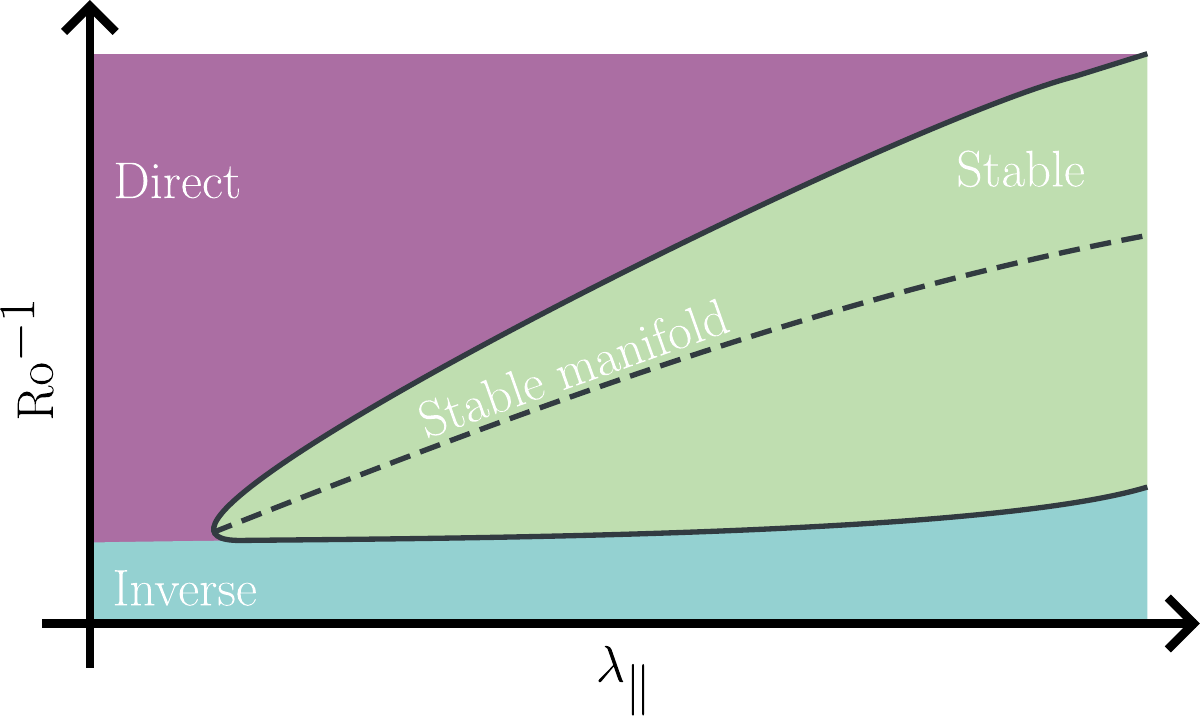}
    \caption{Sketch of the proposed finite-time phase space. A possible stable manifold with zero measure in the limit of very long times is indicated by the dashed line.}
    \label{fig:diagram}
\end{figure}

\section{Conclusions}

We investigated the stability of vortex lattices in classical rotating turbulence using controlled initial conditions that embed clean triangular arrays into fully three-dimensional flows. This approach avoids the defects that arise in spontaneous self-organization of turbulent flows, and allows a systematic exploration of the stability mechanisms and lifetime statistics of the lattices.

A first scan of Rossby numbers and vertical aspect ratios reveals that lattices only appear within a sharply bounded region of parameter space, located between regimes dominated by direct and inverse energy cascades. 
In the absence of an Ekman drag these lattices are not stationary but for small values of $\mathrm{Ro}$ excess inverse transfer leads to energy accumulation and breakup of the lattice by merging, while for large values of $\mathrm{Ro}$ excess direct transfer erodes the lattice through dissipation. These results support the view that lattices are not truly stationary but rather transient states, with the window of appearance shrinking as observation times increase (see a sketch of a proposed phase space in Fig.~\ref{fig:diagram}).

Introducing a dynamically adjusted Ekman drag makes it possible to stabilize the system around prescribed energy levels. This identifies an optimal drag coefficient for which lattice lifetimes are maximized. Deviations from this value---either by overdamping or underdamping---reduce the average lifetime, indicating that the magnitude of the energetic imbalance, rather than its sign, controls the lattice break up.

The use of ensembles further reveals that lattice lifetimes behave as a random variable, with statistics that are approximately memory-less. While two-vortex arrays tend to persist longer than larger lattices, the variance of the lifetime decreases with increasing vortex number, suggesting a thermodynamic-like limit. 

The present results reveal an unexpected richness of dynamics for a very simple system.  The computational costs of performing ensembles of 
runs in a very highly dimensional space allowed us only to scrape tip of the iceberg of this problem leaving many unanswered  questions.
Further investigations are requiered to enlight further this problem.
Finally, carefully designed experiments where vortex-latices can be observed could be a very attractive alternative for future investigations.

\begin{acknowledgments}
The authors acknowledge support from UBACyT Grant No.~20020220300122BA, and from proyecto REMATE of Redes Federales de Alto Impacto, Argentina. JAE thanks Bernardo Español for useful discussions during the writing of the manuscript. 
\end{acknowledgments}

\bibliography{ms}

\begin{thebibliography}{39}%
\makeatletter
\providecommand \@ifxundefined [1]{%
 \@ifx{#1\undefined}
}%
\providecommand \@ifnum [1]{%
 \ifnum #1\expandafter \@firstoftwo
 \else \expandafter \@secondoftwo
 \fi
}%
\providecommand \@ifx [1]{%
 \ifx #1\expandafter \@firstoftwo
 \else \expandafter \@secondoftwo
 \fi
}%
\providecommand \natexlab [1]{#1}%
\providecommand \enquote  [1]{``#1''}%
\providecommand \bibnamefont  [1]{#1}%
\providecommand \bibfnamefont [1]{#1}%
\providecommand \citenamefont [1]{#1}%
\providecommand \href@noop [0]{\@secondoftwo}%
\providecommand \href [0]{\begingroup \@sanitize@url \@href}%
\providecommand \@href[1]{\@@startlink{#1}\@@href}%
\providecommand \@@href[1]{\endgroup#1\@@endlink}%
\providecommand \@sanitize@url [0]{\catcode `\\12\catcode `\$12\catcode `\&12\catcode `\#12\catcode `\^12\catcode `\_12\catcode `\%12\relax}%
\providecommand \@@startlink[1]{}%
\providecommand \@@endlink[0]{}%
\providecommand \url  [0]{\begingroup\@sanitize@url \@url }%
\providecommand \@url [1]{\endgroup\@href {#1}{\urlprefix }}%
\providecommand \urlprefix  [0]{URL }%
\providecommand \Eprint [0]{\href }%
\providecommand \doibase [0]{https://doi.org/}%
\providecommand \selectlanguage [0]{\@gobble}%
\providecommand \bibinfo  [0]{\@secondoftwo}%
\providecommand \bibfield  [0]{\@secondoftwo}%
\providecommand \translation [1]{[#1]}%
\providecommand \BibitemOpen [0]{}%
\providecommand \bibitemStop [0]{}%
\providecommand \bibitemNoStop [0]{.\EOS\space}%
\providecommand \EOS [0]{\spacefactor3000\relax}%
\providecommand \BibitemShut  [1]{\csname bibitem#1\endcsname}%
\let\auto@bib@innerbib\@empty
\bibitem [{\citenamefont {Ouyang}\ and\ \citenamefont {Swinney}(1991)}]{Ouyang_1991}%
  \BibitemOpen
  \bibfield  {author} {\bibinfo {author} {\bibfnamefont {Q.}~\bibnamefont {Ouyang}}\ and\ \bibinfo {author} {\bibfnamefont {H.~L.}\ \bibnamefont {Swinney}},\ }\href@noop {} {\bibfield  {journal} {\bibinfo  {journal} {Nature}\ }\textbf {\bibinfo {volume} {352}},\ \bibinfo {pages} {610} (\bibinfo {year} {1991})}\BibitemShut {NoStop}%
\bibitem [{\citenamefont {Golubitsky}\ \emph {et~al.}(1984)\citenamefont {Golubitsky}, \citenamefont {Swift},\ and\ \citenamefont {Knobloch}}]{Golubitsky_1984}%
  \BibitemOpen
  \bibfield  {author} {\bibinfo {author} {\bibfnamefont {M.}~\bibnamefont {Golubitsky}}, \bibinfo {author} {\bibfnamefont {J.}~\bibnamefont {Swift}},\ and\ \bibinfo {author} {\bibfnamefont {E.}~\bibnamefont {Knobloch}},\ }\href@noop {} {\bibfield  {journal} {\bibinfo  {journal} {Physica D: Nonlinear Phenomena}\ }\textbf {\bibinfo {volume} {10}},\ \bibinfo {pages} {249} (\bibinfo {year} {1984})}\BibitemShut {NoStop}%
\bibitem [{\citenamefont {Herrmann}(2006)}]{Herrmann_2006}%
  \BibitemOpen
  \bibfield  {author} {\bibinfo {author} {\bibfnamefont {H.-J.}\ \bibnamefont {Herrmann}},\ }\href@noop {} {\bibfield  {journal} {\bibinfo  {journal} {Nonlinear Dynamics}\ }\textbf {\bibinfo {volume} {44}},\ \bibinfo {pages} {315} (\bibinfo {year} {2006})}\BibitemShut {NoStop}%
\bibitem [{\citenamefont {Ichihara}\ \emph {et~al.}(2023)\citenamefont {Ichihara}, \citenamefont {Mininni}, \citenamefont {Ravichandran}, \citenamefont {Cimarelli},\ and\ \citenamefont {Vagasky}}]{Ichihara_2023}%
  \BibitemOpen
  \bibfield  {author} {\bibinfo {author} {\bibfnamefont {M.}~\bibnamefont {Ichihara}}, \bibinfo {author} {\bibfnamefont {P.~D.}\ \bibnamefont {Mininni}}, \bibinfo {author} {\bibfnamefont {S.}~\bibnamefont {Ravichandran}}, \bibinfo {author} {\bibfnamefont {C.}~\bibnamefont {Cimarelli}},\ and\ \bibinfo {author} {\bibfnamefont {C.}~\bibnamefont {Vagasky}},\ }\href@noop {} {\bibfield  {journal} {\bibinfo  {journal} {Communications Earth \& Environment}\ }\textbf {\bibinfo {volume} {4}},\ \bibinfo {pages} {417} (\bibinfo {year} {2023})}\BibitemShut {NoStop}%
\bibitem [{\citenamefont {Fernandez-Gonzalez}\ \emph {et~al.}(2024)\citenamefont {Fernandez-Gonzalez}, \citenamefont {Clerc}, \citenamefont {Gonz{\'a}lez-Cort{\'e}s}, \citenamefont {Hidalgo},\ and\ \citenamefont {Vergara}}]{Fernandez_2024}%
  \BibitemOpen
  \bibfield  {author} {\bibinfo {author} {\bibfnamefont {V.}~\bibnamefont {Fernandez-Gonzalez}}, \bibinfo {author} {\bibfnamefont {M.}~\bibnamefont {Clerc}}, \bibinfo {author} {\bibfnamefont {G.}~\bibnamefont {Gonz{\'a}lez-Cort{\'e}s}}, \bibinfo {author} {\bibfnamefont {P.}~\bibnamefont {Hidalgo}},\ and\ \bibinfo {author} {\bibfnamefont {J.}~\bibnamefont {Vergara}},\ }\href@noop {} {\bibfield  {journal} {\bibinfo  {journal} {Reports on Progress in Physics}\ }\textbf {\bibinfo {volume} {87}},\ \bibinfo {pages} {120502} (\bibinfo {year} {2024})}\BibitemShut {NoStop}%
\bibitem [{\citenamefont {Alexakis}\ \emph {et~al.}(2024)\citenamefont {Alexakis}, \citenamefont {Marino}, \citenamefont {Mininni}, \citenamefont {van Kan}, \citenamefont {Foldes},\ and\ \citenamefont {Feraco}}]{Alexakis_2024}%
  \BibitemOpen
  \bibfield  {author} {\bibinfo {author} {\bibfnamefont {A.}~\bibnamefont {Alexakis}}, \bibinfo {author} {\bibfnamefont {R.}~\bibnamefont {Marino}}, \bibinfo {author} {\bibfnamefont {P.~D.}\ \bibnamefont {Mininni}}, \bibinfo {author} {\bibfnamefont {A.}~\bibnamefont {van Kan}}, \bibinfo {author} {\bibfnamefont {R.}~\bibnamefont {Foldes}},\ and\ \bibinfo {author} {\bibfnamefont {F.}~\bibnamefont {Feraco}},\ }\href@noop {} {\bibfield  {journal} {\bibinfo  {journal} {Science}\ }\textbf {\bibinfo {volume} {383}},\ \bibinfo {pages} {1005} (\bibinfo {year} {2024})}\BibitemShut {NoStop}%
\bibitem [{\citenamefont {Ashcroft}\ and\ \citenamefont {Mermin}(1976)}]{Ashcroft}%
  \BibitemOpen
  \bibfield  {author} {\bibinfo {author} {\bibfnamefont {N.}~\bibnamefont {Ashcroft}}\ and\ \bibinfo {author} {\bibfnamefont {N.}~\bibnamefont {Mermin}},\ }\href@noop {} {\emph {\bibinfo {title} {Solid State Physics}}}\ (\bibinfo  {publisher} {Saunders College Publishing},\ \bibinfo {year} {1976})\BibitemShut {NoStop}%
\bibitem [{\citenamefont {Abrikosov}(1957)}]{Abrikosov_1957}%
  \BibitemOpen
  \bibfield  {author} {\bibinfo {author} {\bibfnamefont {A.~A.}\ \bibnamefont {Abrikosov}},\ }\href@noop {} {\bibfield  {journal} {\bibinfo  {journal} {Journal of Physics and Chemistry of Solids}\ }\textbf {\bibinfo {volume} {2}},\ \bibinfo {pages} {199} (\bibinfo {year} {1957})}\BibitemShut {NoStop}%
\bibitem [{\citenamefont {Brandt}(1986)}]{Brandt_1986}%
  \BibitemOpen
  \bibfield  {author} {\bibinfo {author} {\bibfnamefont {E.}~\bibnamefont {Brandt}},\ }\href@noop {} {\bibfield  {journal} {\bibinfo  {journal} {Physical Review B}\ }\textbf {\bibinfo {volume} {34}},\ \bibinfo {pages} {6514} (\bibinfo {year} {1986})}\BibitemShut {NoStop}%
\bibitem [{\citenamefont {Severino}\ \emph {et~al.}(2022)\citenamefont {Severino}, \citenamefont {Mininni}, \citenamefont {Fradkin}, \citenamefont {Bekeris}, \citenamefont {Pasquini}, ,\ and\ \citenamefont {Lozano}}]{Severino_2022}%
  \BibitemOpen
  \bibfield  {author} {\bibinfo {author} {\bibfnamefont {R.~S.}\ \bibnamefont {Severino}}, \bibinfo {author} {\bibfnamefont {P.~D.}\ \bibnamefont {Mininni}}, \bibinfo {author} {\bibfnamefont {E.}~\bibnamefont {Fradkin}}, \bibinfo {author} {\bibfnamefont {V.}~\bibnamefont {Bekeris}}, \bibinfo {author} {\bibfnamefont {G.}~\bibnamefont {Pasquini}}, ,\ and\ \bibinfo {author} {\bibfnamefont {G.~S.}\ \bibnamefont {Lozano}},\ }\href@noop {} {\bibfield  {journal} {\bibinfo  {journal} {Physical Review B}\ }\textbf {\bibinfo {volume} {106}},\ \bibinfo {pages} {094512} (\bibinfo {year} {2022})}\BibitemShut {NoStop}%
\bibitem [{\citenamefont {Fetter}(2009)}]{Fetter_2009}%
  \BibitemOpen
  \bibfield  {author} {\bibinfo {author} {\bibfnamefont {A.~L.}\ \bibnamefont {Fetter}},\ }\href@noop {} {\bibfield  {journal} {\bibinfo  {journal} {Reviews of Modern Physics}\ }\textbf {\bibinfo {volume} {81}},\ \bibinfo {pages} {647} (\bibinfo {year} {2009})}\BibitemShut {NoStop}%
\bibitem [{\citenamefont {Estrada}\ \emph {et~al.}(2022)\citenamefont {Estrada}, \citenamefont {Brachet},\ and\ \citenamefont {Mininni}}]{Amette_Estrada_2022b}%
  \BibitemOpen
  \bibfield  {author} {\bibinfo {author} {\bibfnamefont {J.~A.}\ \bibnamefont {Estrada}}, \bibinfo {author} {\bibfnamefont {M.~E.}\ \bibnamefont {Brachet}},\ and\ \bibinfo {author} {\bibfnamefont {P.~D.}\ \bibnamefont {Mininni}},\ }\href@noop {} {\bibfield  {journal} {\bibinfo  {journal} {{AVS} Quantum Science}\ }\textbf {\bibinfo {volume} {4}},\ \bibinfo {pages} {046201} (\bibinfo {year} {2022})}\BibitemShut {NoStop}%
\bibitem [{\citenamefont {Amette~Estrada}\ \emph {et~al.}(2025)\citenamefont {Amette~Estrada}, \citenamefont {Brachet},\ and\ \citenamefont {Mininni}}]{Amette_2025}%
  \BibitemOpen
  \bibfield  {author} {\bibinfo {author} {\bibfnamefont {J.}~\bibnamefont {Amette~Estrada}}, \bibinfo {author} {\bibfnamefont {M.~E.}\ \bibnamefont {Brachet}},\ and\ \bibinfo {author} {\bibfnamefont {P.~D.}\ \bibnamefont {Mininni}},\ }\href@noop {} {\bibfield  {journal} {\bibinfo  {journal} {Physical Review A}\ }\textbf {\bibinfo {volume} {111}},\ \bibinfo {pages} {023304} (\bibinfo {year} {2025})}\BibitemShut {NoStop}%
\bibitem [{\citenamefont {de~Wit}\ \emph {et~al.}(2024)\citenamefont {de~Wit}, \citenamefont {Fruchart}, \citenamefont {Khain}, \citenamefont {Toschi},\ and\ \citenamefont {Vitelli}}]{Dewit_2024}%
  \BibitemOpen
  \bibfield  {author} {\bibinfo {author} {\bibfnamefont {X.~M.}\ \bibnamefont {de~Wit}}, \bibinfo {author} {\bibfnamefont {M.}~\bibnamefont {Fruchart}}, \bibinfo {author} {\bibfnamefont {T.}~\bibnamefont {Khain}}, \bibinfo {author} {\bibfnamefont {F.}~\bibnamefont {Toschi}},\ and\ \bibinfo {author} {\bibfnamefont {V.}~\bibnamefont {Vitelli}},\ }\href@noop {} {\bibfield  {journal} {\bibinfo  {journal} {Nature}\ }\textbf {\bibinfo {volume} {627}},\ \bibinfo {pages} {515} (\bibinfo {year} {2024})}\BibitemShut {NoStop}%
\bibitem [{\citenamefont {Floyd}\ \emph {et~al.}(2024)\citenamefont {Floyd}, \citenamefont {Dinner},\ and\ \citenamefont {Vaikuntanathan}}]{Floyd_2024}%
  \BibitemOpen
  \bibfield  {author} {\bibinfo {author} {\bibfnamefont {C.}~\bibnamefont {Floyd}}, \bibinfo {author} {\bibfnamefont {A.~R.}\ \bibnamefont {Dinner}},\ and\ \bibinfo {author} {\bibfnamefont {S.}~\bibnamefont {Vaikuntanathan}},\ }\href {https://doi.org/10.1103/PhysRevResearch.6.033100} {\bibfield  {journal} {\bibinfo  {journal} {Phys. Rev. Res.}\ }\textbf {\bibinfo {volume} {6}},\ \bibinfo {pages} {033100} (\bibinfo {year} {2024})}\BibitemShut {NoStop}%
\bibitem [{\citenamefont {Boubnov}\ and\ \citenamefont {Golitsyn}(1986)}]{Boubnov_1986}%
  \BibitemOpen
  \bibfield  {author} {\bibinfo {author} {\bibfnamefont {B.~M.}\ \bibnamefont {Boubnov}}\ and\ \bibinfo {author} {\bibfnamefont {G.~S.}\ \bibnamefont {Golitsyn}},\ }\href {https://doi.org/10.1017/S002211208600294X} {\bibfield  {journal} {\bibinfo  {journal} {Journal of Fluid Mechanics}\ }\textbf {\bibinfo {volume} {167}},\ \bibinfo {pages} {503} (\bibinfo {year} {1986})}\BibitemShut {NoStop}%
\bibitem [{\citenamefont {Boubnov}\ and\ \citenamefont {Golitsyn}(1990)}]{Boubnov_1990}%
  \BibitemOpen
  \bibfield  {author} {\bibinfo {author} {\bibfnamefont {B.~M.}\ \bibnamefont {Boubnov}}\ and\ \bibinfo {author} {\bibfnamefont {G.~S.}\ \bibnamefont {Golitsyn}},\ }\href {https://doi.org/10.1017/S0022112090002920} {\bibfield  {journal} {\bibinfo  {journal} {Journal of Fluid Mechanics}\ }\textbf {\bibinfo {volume} {219}},\ \bibinfo {pages} {215} (\bibinfo {year} {1990})}\BibitemShut {NoStop}%
\bibitem [{\citenamefont {van Kan}\ \emph {et~al.}(2024)\citenamefont {van Kan}, \citenamefont {Favier}, \citenamefont {Julien},\ and\ \citenamefont {Knobloch}}]{AvK_2024b}%
  \BibitemOpen
  \bibfield  {author} {\bibinfo {author} {\bibfnamefont {A.}~\bibnamefont {van Kan}}, \bibinfo {author} {\bibfnamefont {B.}~\bibnamefont {Favier}}, \bibinfo {author} {\bibfnamefont {K.}~\bibnamefont {Julien}},\ and\ \bibinfo {author} {\bibfnamefont {E.}~\bibnamefont {Knobloch}},\ }\href@noop {} {\bibfield  {journal} {\bibinfo  {journal} {Journal of Fluid Mechanics}\ }\textbf {\bibinfo {volume} {984}} (\bibinfo {year} {2024})}\BibitemShut {NoStop}%
\bibitem [{\citenamefont {Clark Di~Leoni}\ \emph {et~al.}(2020)\citenamefont {Clark Di~Leoni}, \citenamefont {Alexakis}, \citenamefont {Biferale},\ and\ \citenamefont {Buzzicotti}}]{ClarkDiLeoni2020}%
  \BibitemOpen
  \bibfield  {author} {\bibinfo {author} {\bibfnamefont {P.}~\bibnamefont {Clark Di~Leoni}}, \bibinfo {author} {\bibfnamefont {A.}~\bibnamefont {Alexakis}}, \bibinfo {author} {\bibfnamefont {L.}~\bibnamefont {Biferale}},\ and\ \bibinfo {author} {\bibfnamefont {M.}~\bibnamefont {Buzzicotti}},\ }\bibfield  {journal} {\bibinfo  {journal} {Physical Review Fluids}\ }\textbf {\bibinfo {volume} {5}},\ \href {https://doi.org/10.1103/physrevfluids.5.104603} {10.1103/physrevfluids.5.104603} (\bibinfo {year} {2020})\BibitemShut {NoStop}%
\bibitem [{\citenamefont {Marchetti}\ and\ \citenamefont {Mininni}(2025)}]{Marchetti_2025}%
  \BibitemOpen
  \bibfield  {author} {\bibinfo {author} {\bibfnamefont {G.}~\bibnamefont {Marchetti}}\ and\ \bibinfo {author} {\bibfnamefont {P.~D.}\ \bibnamefont {Mininni}},\ }\href@noop {} {\bibfield  {journal} {\bibinfo  {journal} {Phys. Rev. Fluids}\ }\textbf {\bibinfo {volume} {10}},\ \bibinfo {pages} {084603} (\bibinfo {year} {2025})}\BibitemShut {NoStop}%
\bibitem [{\citenamefont {Adriani}\ \emph {et~al.}(2018)\citenamefont {Adriani}, \citenamefont {Mura}, \citenamefont {Orton},\ and\ \citenamefont {et~al.}}]{Adriani_2018}%
  \BibitemOpen
  \bibfield  {author} {\bibinfo {author} {\bibfnamefont {A.}~\bibnamefont {Adriani}}, \bibinfo {author} {\bibfnamefont {A.}~\bibnamefont {Mura}}, \bibinfo {author} {\bibfnamefont {G.}~\bibnamefont {Orton}},\ and\ \bibinfo {author} {\bibnamefont {et~al.}},\ }\href@noop {} {\bibfield  {journal} {\bibinfo  {journal} {Nature}\ }\textbf {\bibinfo {volume} {555}},\ \bibinfo {pages} {216–219} (\bibinfo {year} {2018})}\BibitemShut {NoStop}%
\bibitem [{\citenamefont {Moisy}\ \emph {et~al.}(2011)\citenamefont {Moisy}, \citenamefont {Morize}, \citenamefont {Rabaud},\ and\ \citenamefont {Sommeria}}]{Moisy_2010}%
  \BibitemOpen
  \bibfield  {author} {\bibinfo {author} {\bibfnamefont {F.}~\bibnamefont {Moisy}}, \bibinfo {author} {\bibfnamefont {C.}~\bibnamefont {Morize}}, \bibinfo {author} {\bibfnamefont {M.}~\bibnamefont {Rabaud}},\ and\ \bibinfo {author} {\bibfnamefont {J.}~\bibnamefont {Sommeria}},\ }\href@noop {} {\bibfield  {journal} {\bibinfo  {journal} {Journal of Fluid Mechanics}\ }\textbf {\bibinfo {volume} {666}},\ \bibinfo {pages} {5–35} (\bibinfo {year} {2011})}\BibitemShut {NoStop}%
\bibitem [{\citenamefont {Gallet}\ \emph {et~al.}(2014)\citenamefont {Gallet}, \citenamefont {Campagne}, \citenamefont {Cortet},\ and\ \citenamefont {Moisy}}]{Gallet_2014}%
  \BibitemOpen
  \bibfield  {author} {\bibinfo {author} {\bibfnamefont {B.}~\bibnamefont {Gallet}}, \bibinfo {author} {\bibfnamefont {A.}~\bibnamefont {Campagne}}, \bibinfo {author} {\bibfnamefont {P.-P.}\ \bibnamefont {Cortet}},\ and\ \bibinfo {author} {\bibfnamefont {F.}~\bibnamefont {Moisy}},\ }\href@noop {} {\bibfield  {journal} {\bibinfo  {journal} {Physics of Fluids}\ }\textbf {\bibinfo {volume} {26}},\ \bibinfo {pages} {035108} (\bibinfo {year} {2014})}\BibitemShut {NoStop}%
\bibitem [{\citenamefont {Aref}\ and\ \citenamefont {Vainchtein}(1998)}]{aref1998point}%
  \BibitemOpen
  \bibfield  {author} {\bibinfo {author} {\bibfnamefont {H.}~\bibnamefont {Aref}}\ and\ \bibinfo {author} {\bibfnamefont {D.~L.}\ \bibnamefont {Vainchtein}},\ }\href@noop {} {\bibfield  {journal} {\bibinfo  {journal} {Nature}\ }\textbf {\bibinfo {volume} {392}},\ \bibinfo {pages} {769} (\bibinfo {year} {1998})}\BibitemShut {NoStop}%
\bibitem [{\citenamefont {Cambon}\ \emph {et~al.}(2004)\citenamefont {Cambon}, \citenamefont {Rubinstein},\ and\ \citenamefont {Godeferd}}]{Cambon_2004}%
  \BibitemOpen
  \bibfield  {author} {\bibinfo {author} {\bibfnamefont {C.}~\bibnamefont {Cambon}}, \bibinfo {author} {\bibfnamefont {R.}~\bibnamefont {Rubinstein}},\ and\ \bibinfo {author} {\bibfnamefont {F.~S.}\ \bibnamefont {Godeferd}},\ }\href@noop {} {\bibfield  {journal} {\bibinfo  {journal} {New Journal of Physics}\ }\textbf {\bibinfo {volume} {6}},\ \bibinfo {pages} {73} (\bibinfo {year} {2004})}\BibitemShut {NoStop}%
\bibitem [{\citenamefont {Sen}\ \emph {et~al.}(2012)\citenamefont {Sen}, \citenamefont {Mininni}, \citenamefont {Rosenberg},\ and\ \citenamefont {Pouquet}}]{Sen_2012}%
  \BibitemOpen
  \bibfield  {author} {\bibinfo {author} {\bibfnamefont {A.}~\bibnamefont {Sen}}, \bibinfo {author} {\bibfnamefont {P.~D.}\ \bibnamefont {Mininni}}, \bibinfo {author} {\bibfnamefont {D.}~\bibnamefont {Rosenberg}},\ and\ \bibinfo {author} {\bibfnamefont {A.}~\bibnamefont {Pouquet}},\ }\href@noop {} {\bibfield  {journal} {\bibinfo  {journal} {Physical Review E}\ }\textbf {\bibinfo {volume} {86}} (\bibinfo {year} {2012})}\BibitemShut {NoStop}%
\bibitem [{\citenamefont {Deusebio}\ \emph {et~al.}(2014)\citenamefont {Deusebio}, \citenamefont {Boffetta}, \citenamefont {Lindborg},\ and\ \citenamefont {Musacchio}}]{Deusebio_2014}%
  \BibitemOpen
  \bibfield  {author} {\bibinfo {author} {\bibfnamefont {E.}~\bibnamefont {Deusebio}}, \bibinfo {author} {\bibfnamefont {G.}~\bibnamefont {Boffetta}}, \bibinfo {author} {\bibfnamefont {E.}~\bibnamefont {Lindborg}},\ and\ \bibinfo {author} {\bibfnamefont {S.}~\bibnamefont {Musacchio}},\ }\href@noop {} {\bibfield  {journal} {\bibinfo  {journal} {Physical Review E}\ }\textbf {\bibinfo {volume} {90}},\ \bibinfo {pages} {023005} (\bibinfo {year} {2014})}\BibitemShut {NoStop}%
\bibitem [{\citenamefont {van Kan}\ and\ \citenamefont {Alexakis}(2020)}]{vanKan2020}%
  \BibitemOpen
  \bibfield  {author} {\bibinfo {author} {\bibfnamefont {A.}~\bibnamefont {van Kan}}\ and\ \bibinfo {author} {\bibfnamefont {A.}~\bibnamefont {Alexakis}},\ }\bibfield  {journal} {\bibinfo  {journal} {Journal of Fluid Mechanics}\ }\textbf {\bibinfo {volume} {899}},\ \href {https://doi.org/10.1017/jfm.2020.443} {10.1017/jfm.2020.443} (\bibinfo {year} {2020})\BibitemShut {NoStop}%
\bibitem [{\citenamefont {Alexakis}\ and\ \citenamefont {Biferale}(2018)}]{Alexakis2018}%
  \BibitemOpen
  \bibfield  {author} {\bibinfo {author} {\bibfnamefont {A.}~\bibnamefont {Alexakis}}\ and\ \bibinfo {author} {\bibfnamefont {L.}~\bibnamefont {Biferale}},\ }\href {https://doi.org/10.1016/j.physrep.2018.08.001} {\bibfield  {journal} {\bibinfo  {journal} {Physics Reports}\ }\textbf {\bibinfo {volume} {767–769}},\ \bibinfo {pages} {1–101} (\bibinfo {year} {2018})}\BibitemShut {NoStop}%
\bibitem [{\citenamefont {Celani}\ \emph {et~al.}(2010)\citenamefont {Celani}, \citenamefont {Musacchio},\ and\ \citenamefont {Vincenzi}}]{Celani_2010}%
  \BibitemOpen
  \bibfield  {author} {\bibinfo {author} {\bibfnamefont {A.}~\bibnamefont {Celani}}, \bibinfo {author} {\bibfnamefont {S.}~\bibnamefont {Musacchio}},\ and\ \bibinfo {author} {\bibfnamefont {D.}~\bibnamefont {Vincenzi}},\ }\href@noop {} {\bibfield  {journal} {\bibinfo  {journal} {Physical Review Letters}\ }\textbf {\bibinfo {volume} {104}},\ \bibinfo {pages} {184506} (\bibinfo {year} {2010})}\BibitemShut {NoStop}%
\bibitem [{\citenamefont {Benavides}\ and\ \citenamefont {Alexakis}(2017)}]{Benavides_2017}%
  \BibitemOpen
  \bibfield  {author} {\bibinfo {author} {\bibfnamefont {S.~J.}\ \bibnamefont {Benavides}}\ and\ \bibinfo {author} {\bibfnamefont {A.}~\bibnamefont {Alexakis}},\ }\href@noop {} {\bibfield  {journal} {\bibinfo  {journal} {Journal of Fluid Mechanics}\ }\textbf {\bibinfo {volume} {822}},\ \bibinfo {pages} {364} (\bibinfo {year} {2017})}\BibitemShut {NoStop}%
\bibitem [{\citenamefont {Aref}\ \emph {et~al.}(2003)\citenamefont {Aref}, \citenamefont {Newton}, \citenamefont {Stremler}, \citenamefont {Tokieda},\ and\ \citenamefont {Vainchtein}}]{Aref_2003}%
  \BibitemOpen
  \bibfield  {author} {\bibinfo {author} {\bibfnamefont {H.}~\bibnamefont {Aref}}, \bibinfo {author} {\bibfnamefont {P.}~\bibnamefont {Newton}}, \bibinfo {author} {\bibfnamefont {M.}~\bibnamefont {Stremler}}, \bibinfo {author} {\bibfnamefont {T.}~\bibnamefont {Tokieda}},\ and\ \bibinfo {author} {\bibfnamefont {D.}~\bibnamefont {Vainchtein}},\ }\href {https://doi.org/10.1016/S0065-2156(02)39001-X} {\bibfield  {journal} {\bibinfo  {journal} {Advances in Applied Mechanics}\ }\textbf {\bibinfo {volume} {39}},\ \bibinfo {pages} {1} (\bibinfo {year} {2003})}\BibitemShut {NoStop}%
\bibitem [{\citenamefont {Mininni}\ \emph {et~al.}(2011)\citenamefont {Mininni}, \citenamefont {Rosenberg}, \citenamefont {Reddy},\ and\ \citenamefont {Pouquet}}]{Mininni2011}%
  \BibitemOpen
  \bibfield  {author} {\bibinfo {author} {\bibfnamefont {P.~D.}\ \bibnamefont {Mininni}}, \bibinfo {author} {\bibfnamefont {D.}~\bibnamefont {Rosenberg}}, \bibinfo {author} {\bibfnamefont {R.}~\bibnamefont {Reddy}},\ and\ \bibinfo {author} {\bibfnamefont {A.}~\bibnamefont {Pouquet}},\ }\href {https://doi.org/10.1016/j.parco.2011.05.004} {\bibfield  {journal} {\bibinfo  {journal} {Parallel Computing}\ }\textbf {\bibinfo {volume} {37}},\ \bibinfo {pages} {316} (\bibinfo {year} {2011})}\BibitemShut {NoStop}%
\bibitem [{\citenamefont {Rosenberg}\ \emph {et~al.}(2020)\citenamefont {Rosenberg}, \citenamefont {Mininni}, \citenamefont {Reddy},\ and\ \citenamefont {Pouquet}}]{Rosenberg_2020}%
  \BibitemOpen
  \bibfield  {author} {\bibinfo {author} {\bibfnamefont {D.}~\bibnamefont {Rosenberg}}, \bibinfo {author} {\bibfnamefont {P.~D.}\ \bibnamefont {Mininni}}, \bibinfo {author} {\bibfnamefont {R.}~\bibnamefont {Reddy}},\ and\ \bibinfo {author} {\bibfnamefont {A.}~\bibnamefont {Pouquet}},\ }\href@noop {} {\bibfield  {journal} {\bibinfo  {journal} {Atmosphere}\ }\textbf {\bibinfo {volume} {11}},\ \bibinfo {pages} {178} (\bibinfo {year} {2020})}\BibitemShut {NoStop}%
\bibitem [{\citenamefont {Seshasayanan}\ and\ \citenamefont {Gallet}(2020)}]{seshasayanan2020onset}%
  \BibitemOpen
  \bibfield  {author} {\bibinfo {author} {\bibfnamefont {K.}~\bibnamefont {Seshasayanan}}\ and\ \bibinfo {author} {\bibfnamefont {B.}~\bibnamefont {Gallet}},\ }\href@noop {} {\bibfield  {journal} {\bibinfo  {journal} {Journal of Fluid Mechanics}\ }\textbf {\bibinfo {volume} {901}},\ \bibinfo {pages} {R5} (\bibinfo {year} {2020})}\BibitemShut {NoStop}%
\bibitem [{\citenamefont {Lohani}\ \emph {et~al.}(2024)\citenamefont {Lohani}, \citenamefont {Nayak},\ and\ \citenamefont {Seshasayanan}}]{lohani2024effect}%
  \BibitemOpen
  \bibfield  {author} {\bibinfo {author} {\bibfnamefont {C.~S.}\ \bibnamefont {Lohani}}, \bibinfo {author} {\bibfnamefont {S.~K.}\ \bibnamefont {Nayak}},\ and\ \bibinfo {author} {\bibfnamefont {K.}~\bibnamefont {Seshasayanan}},\ }\href@noop {} {\bibfield  {journal} {\bibinfo  {journal} {Physical Review Fluids}\ }\textbf {\bibinfo {volume} {9}},\ \bibinfo {pages} {034604} (\bibinfo {year} {2024})}\BibitemShut {NoStop}%
\bibitem [{\citenamefont {Das}\ \emph {et~al.}(2025)\citenamefont {Das}, \citenamefont {Sharma}, \citenamefont {Ranjan},\ and\ \citenamefont {Verma}}]{Das_2025}%
  \BibitemOpen
  \bibfield  {author} {\bibinfo {author} {\bibfnamefont {A.}~\bibnamefont {Das}}, \bibinfo {author} {\bibfnamefont {M.}~\bibnamefont {Sharma}}, \bibinfo {author} {\bibfnamefont {A.}~\bibnamefont {Ranjan}},\ and\ \bibinfo {author} {\bibfnamefont {M.~K.}\ \bibnamefont {Verma}},\ }\href {https://doi.org/10.1103/5kw9-kz43} {\bibfield  {journal} {\bibinfo  {journal} {Phys. Rev. Fluids}\ }\textbf {\bibinfo {volume} {10}},\ \bibinfo {pages} {084803} (\bibinfo {year} {2025})}\BibitemShut {NoStop}%
\bibitem [{\citenamefont {Yarmchuk}\ \emph {et~al.}(1979)\citenamefont {Yarmchuk}, \citenamefont {Gordon},\ and\ \citenamefont {Packard}}]{Yarmchuk79}%
  \BibitemOpen
  \bibfield  {author} {\bibinfo {author} {\bibfnamefont {E.~J.}\ \bibnamefont {Yarmchuk}}, \bibinfo {author} {\bibfnamefont {M.~J.~V.}\ \bibnamefont {Gordon}},\ and\ \bibinfo {author} {\bibfnamefont {R.~E.}\ \bibnamefont {Packard}},\ }\href {https://doi.org/10.1103/PhysRevLett.43.214} {\bibfield  {journal} {\bibinfo  {journal} {Phys. Rev. Lett.}\ }\textbf {\bibinfo {volume} {43}},\ \bibinfo {pages} {214} (\bibinfo {year} {1979})}\BibitemShut {NoStop}%
\bibitem [{\citenamefont {Kragset}\ \emph {et~al.}(2006)\citenamefont {Kragset}, \citenamefont {Babaev},\ and\ \citenamefont {Sudbø}}]{Kragset2006}%
  \BibitemOpen
  \bibfield  {author} {\bibinfo {author} {\bibfnamefont {S.}~\bibnamefont {Kragset}}, \bibinfo {author} {\bibfnamefont {E.}~\bibnamefont {Babaev}},\ and\ \bibinfo {author} {\bibfnamefont {A.}~\bibnamefont {Sudbø}},\ }\bibfield  {journal} {\bibinfo  {journal} {Physical Review Letters}\ }\textbf {\bibinfo {volume} {97}},\ \href {https://doi.org/10.1103/physrevlett.97.170403} {10.1103/physrevlett.97.170403} (\bibinfo {year} {2006})\BibitemShut {NoStop}%
\end{thebibliography}%

\end{document}